\documentclass[aps, twocolumn, showpacs]{revtex4-2}
\usepackage{amsmath}
\begin{document}
\title {Comments on "A clarification on prevailing misconceptions in unimodular gravity"}

\author{S. C. Tiwari \\
Department of Physics, Institute of Science, Banaras Hindu University, Varanasi 221005, and \\ Institute of Natural Philosophy \\
Varanasi India\\ Email address: $vns\_sctiwari@yahoo.com$ \\}
\begin{abstract}
In this comment on arXiv:2308.07360 it is pointed out that though the authors raise an important question on the prevailing presentations of the unimodular gravity their own presentation is incomplete as well as confusing equating assumptions to misconceptions. We focus on the notion of diffeomorphism invariance, general covariance and the conservation/nonconservation of the energy-momentum tensor in this short note to delineate the fundamental issues in unimodular gravity. The present discussion is intended to offer a wider and deeper perspective on the objective set by the authors in their paper.
\end{abstract}

\maketitle

{\bf I:} First two sections in \cite{1} explain the main issues that the authors seek to clarify on the unimodular gravity (UG). Essentially there are three points of concern: (i) non-dynamical 4-volume element in UG does not imply the loss of general covariance, (ii) the prescription to decide the dynamical or non-dynamical nature of an object, and (iii) the conservation law for the energy-momentum tensor in both general relativity (GR) and UG. In the beginning of the article authors state:"The purpose of this manuscript is to attempt to clarify some of the confusions and misconceptions that appear in several of the works mentioned." It is not clear which of the 31 references cited by them have confusions and misconceptions; making such a sweeping statement has not been supported by explicit examples of the misconceptions. I think there do exist problems in the presentation of UG in the literature. However, rather than misconceptions the main questions relate with the lack of understanding on Einstein's three papers linked with the cosmological constant in different ways, and the assumptions on the energy-momentum conservation that could be debated/modified in both GR and UG \cite{2}.  Such presentations in the literature do not belong to misconceptions: I will  discuss Finkelstein et al paper \cite{3} i. e.  reference 5 in \cite{1} to explain this point in connection with the law of energy conservation. 

Regarding the first two points it has to be emphasized that Einstein himself asserted rightly that the assumption $\sqrt{-g} =1$ did not imply abandonement of general covariance \cite{2}. Weinberg \cite{4} in Section VII of his paper notes that $\sqrt{-g}$  is non-dynamical and the theory (UG) is generally covariant. In one of the earliest papers on UG in a particle physics oriented approach \cite{5} in Section 5 authors explain that any solution could be covered with coordinate patches where  $\sqrt{-g} =1$. Quite often the 4-volume preserving symmetry and transverse diffeomorphisms  are used in UG; however they are not identical, see \cite{6} for a pedagogical account. Einstein and Eddington did appreciate the importance of physical interpretation of the invariance of the 4-volume element, see discussion in \cite{2}. A nontrivial distinction between GR and UG will arise if we free UG from the cosmological constant paradigm and consider the geometry of paths in an affine manifold \cite{2}. In fact, a recent paper seems to appreciate this aspect \cite{7}. What is the cosmological constant paradigm? We refer to \cite{2} and make additional remarks here.  UG in Weinberg's review appears just in one small section (VII), and footnote on page 12 \cite{4} shows that he had hardly any appreciation of Einstein's work. In a review on the problem of the cosmological constant\cite{8} Ng in Section 4.9 notes that, 'But apparently he (Einstein) did not realize the connection between the unimodular condition and the cosmological constant being an integration constant. Unaware of Einstein's work in this area, quite a few groups of people, including van Dam, van der Bij, and myself, have (re)studied the unimodular or related theories'. I think Ng is partially correct, and it is historically wrong to attribute UG to Einstein, and there arise conceptual flaws if Einstein's work is not studied thoroughly \cite{2}.

The role of dynamical/non-dynamical field variables, field equations and the action integral in a physical theory have intricate and intriguing aspects. Some of them are nicely discussed in \cite{1}. For a wider and deeper perpective we add some remarks. General covariance in Maxwell theory and the constitutive relations has been discussed comprehensively by Post in a monograph \cite{9}. In my own work \cite{10} on local duality symmetry it has been found that Maxwell equations could be generalized possessing local duality invariance, however, unusual pseudo-vector Lagrangian density had to be used for the variational formulation of the action integral. In the second example, there exists unsolved problem of constructing the action functional for the Einstein-Weyl equation. In \cite{11} an attempt has been made to derive the restricted class of the Einstein-Weyl equation from the variational principle. 

{\bf II:} It is known that the conservation of matter energy-momentum tensor $T_{\mu\nu}$ played key role in the developments leading to the Einstein field equation. However, since the beginning, the physical import of the local energy conservation has been debatable; Einstein himself introduced a pseudo-tensor to address this question in GR. A recent paper \cite{12} re-visits covariant divergence law
\begin{equation}
T^{\mu\nu} _{~:\nu} =0
\end{equation}
using Noether's (second) theorem, and compares this formalism with earlier works. Equation (51) in \cite{12} leads to Eq.(1) only if the cosmological constant is assumed to be a constant. Note that the first proposition modifying the law (1) in the context of UG was made in \cite{13}.  Finkelstein et al \cite{3} motivated by this work \cite{13} re-examined the whole issue maintaining that the cosmological constant was an integration constant. However, we pointed out \cite{14} that the principal conclusion in \cite{3} was an assumption not a consequence of the theory. However, there is no misconception here. Authors \cite{1} cite Finkelstein et al paper but fail to recognize its origin and import: it is straightforward to verify that the discussion beginning from Equation (2.13) and ending with the statement " Nevertheless, in principle, one might choose not to adopt such an assumption, in which case, the important fact is that UG generically admits a violation of the energy-momentum conservation, as long as the 1-form J is closed. It is for this last case that UG introduces deviations from GR" in their paper \cite{1} were already discussed in my paper \cite{13}.

In the context of the statement in \cite{1} that " The first point to note is that the only notion of integration over an n-dimensional manifold that is mathematically well defined (given the very notion of what a differentiable manifold is), is the integral of an n-form" it would be useful to recall interesting remarks by Regge \cite{15}. In the language of differential forms 'the metric tensor thus became an auxiliary quantity, Christoffel symbols disappeared from the scene and it became quite easy to deal with spinors'. The metric tensor in terms of tetrad $e^a_\mu$ is 
\begin{equation}
g_{\mu\nu} = e^a_\mu \otimes e^b_\nu \eta_{\mu\nu}
\end{equation}
The connection in terms of 1-form $e^a = e^a_\mu dx^\mu$ is defined by
\begin{equation}
T^a = de^a +\omega^a_{~b} \wedge e^b =0
\end{equation}
and the Ricci curvature is
\begin{equation}
R^{ab} =d\omega^{ab} + \omega^a \wedge \omega^b
\end{equation}
The Einstein-Hilbert action
\begin{equation}
\int R^{ab} \wedge e^c \wedge e^d ~\epsilon_{abcd}
\end{equation}
using variational principle gives the Einstein field equation
\begin{equation}
R^{ab} \wedge e^c ~\epsilon_{abcd} =0
\end{equation}
Now, the crucial point is that the action (5) is not invariant under the general Poincare transformations that include space-time translations. Regge considers it an embarassing notion and looks unfavorably on the amended gauge transformations to remedy it.

A recent paper \cite{16} considers first order formalism of Palatini-type and derives trace-free Einstein field equations from a diffeomorphism invariant BF-type action. Solutions given by expressions (8) and (9) in this paper are shown to correspond to trace-free Einstein field equations and GR equations respectively. It would seem interesting to examine if the expression (8) is related with the  condition for the  equi-projective geometry or the vanishing of $\Gamma^\sigma_{~\mu \sigma}$ \cite{2}. Note that in the Palatini action the affine connection is treated as an independent field variable. 

{\bf III:~} In conclusion, I believe there is a need to clarify fundamental issues in UG, and the present note would be useful to achieve this objective set out in \cite{1}.

{\bf Acknowledgment}

I thank Professor Sinya Aoki for a correspondence.

\end{document}